\documentstyle[12pt,tighten,aps,german,a4]{revtex}

\begin{document}
\title{Lyapunov Exponents from Kinetic Theory for a Dilute, Field-driven
Lorentz Gas}
\author{H. van Beijeren[*], J. R. Dorfman[**], E. G. D. Cohen[{$\dagger$}],
H. A. Posch[$\ddagger$], and Ch. Dellago[$\ddagger$]}
\address{* Institute for Theoretical Physics, University of Utrecht, 
Postbus 80006, 3508 TA
Utrecht, The Netherlands;\\ $**$ I.P.S.T. and Department of Physics, 
University of
Maryland, College Park, MD, 20742, USA;\\ $\dagger$ The Rockefeller 
University , 1230
York Ave., New York, NY, 10021, USA;\\ $\ddagger$ Institut f\"{u}r 
Experimentalphysik,
Universit\"{a}t Wien, Boltzmanngasse 5, A-1090 Wien, Austria}
\date{\today}
\maketitle
\begin{abstract}
Positive and negative Lyapunov exponents for a dilute, random, 
two-dimensional Lorentz gas in an applied field,
$\vec{E}$, in a steady state at constant energy are computed to
order $E^{2}$. The results are:
$\lambda_{\pm}=\lambda_{\pm}^{0}-a_{\pm}(qE/mv)^{2}t_{0}$ where
$\lambda_{\pm}^{0}$ are the exponents for the field-free Lorentz gas,
$a_{+}=11/48,  a_{-}=7/48$, $t_{0}$ is the mean free time between
collisions, $q$ is the charge, $m$ the mass and $v$ is the 
speed of the particle. 
The calculation is
based on an extended Boltzmann equation in which a radius of
curvature, characterizing the separation of
two nearby trajectories, is one of the variables in the
distribution function. The analytical results are in excellent
agreement with computer simulations. These simulations provide
additional evidence for logarithmic terms in the density expansion 
of the diffusion
coefficient.
\end{abstract}

     One of the outstanding problems in transport theory is to relate
macroscopic quantities such as transport coefficients to microscopic
quantities that characterize the chaotic dynamics of the system. This
chaotic dynamics is responsible for the stochastic-like behavior which
leads to normal hydrodynamic processes taking place
in fluids. One approach is to consider the system as a
Hamiltonian system, obeying classical mechanics, and to apply the
escape-rate formalism which can be used to express transport coefficients in
terms of Lyapunov exponents and Kolmogorov-Sinai entropies for 
trajectories in phase space
on an appropriate fractal repeller \cite{gn89}. Another approach 
to this problem is to
consider driven, thermostatted systems, in a nonequilibrium steady
state where the system is subjected
to an applied force as well as to a 
``Gaussian thermostat'',  
which allows the system to reach a steady state by removing the
heat produced by the applied force \cite{em,hoov}. Here one can
relate the transport coefficients to the change in the sum of all, or
in some cases, a pair \cite{em}, of the
Lyapunov exponents in the steady state.

      The system considered
here is a dilute, thermostatted two-dimensional Lorentz gas, where a
particle with charge $q$ and mass $m$ moves in an infinite random 
array of fixed hard disk
scatterers of radius $a$ and density $n, na^{2}\ll 1$, subject to a
constant, uniform, external field $\vec{E}$ as well as to a frictional
thermostat which maintains the speed of the particle at a constant value
$v$. The system approaches a non-equilibrium steady state,
characterized by an attractor in phase space. In this steady state, the 
following relation holds for the macroscopic
diffusion coefficient, $D$, for the particle (or, equivalently, the conductivity
$\sigma$ related to $D$ by an Einstein formula $D=m(v/q)^{2}\sigma$ ) and the Lyapunov exponents
$\lambda_{+}(\varepsilon)$ and $\lambda_{-}(\varepsilon)$ \cite{em,ecsl93}:
$D =  - \lim_{\varepsilon\rightarrow 0}[ v^2 (\lambda _{+}(\varepsilon)+\lambda
_{-}(\varepsilon))]\varepsilon^{-2}$,
where $\varepsilon = (qE/vm)$.
       Here we report the first theoretical calculation of both $\lambda
_{+}(\varepsilon)$ and $%
\lambda _{-}(\varepsilon)$, for this system,
using kinetic theory. We also present a comparison with extensive
computer simulations, as well as with the general relation,
given above. We apply a
method based on the extended Lorentz-Boltzmann (LB) equation developed by van
Beijeren and Dorfman for computing the Lyapunov exponents \cite{vbd95,lvbd}.

      The motion of the particle is described by a
position vector $\vec{r}$, and  velocity $\vec{v}$. Between collisions with
the disks, the (non-Hamiltonian)
equations of motion of the particle in the field and with an energy
conserving Gaussian thermostat are 
\begin{eqnarray}
\dot{x} =v_x=v \cos \theta;\;\;\;\dot{y} = v_{y} = v \sin \theta \nonumber \\
\dot{v}_x =v\varepsilon-\alpha v_x;\;\;\;\dot{v}_y =-\alpha v_y.  \label{2} 
\end{eqnarray}
Here $\theta $ is the angle the particle's velocity makes with the applied field, 
in the $x$-direction, and $\alpha = v_x\varepsilon/v$
represents the strength of the frictional force provided by the Gaussian
thermostat \cite{em}, determined by the condition that the
kinetic energy remain constant.

The instantaneous change in velocity of the moving 
particle upon collisions is
$\vec{v'}-\vec{v}=-2\hat{k}(\vec{v}\cdot \hat{k})$, 
where $\vec{v'}$ is its velocity after
collision, and $\hat{k}$ is the unit vector
in the direction from the center
of the scatterer to the point of impact.

To determine
the positive Lyapunov exponent, $\lambda_{+}$ we first 
consider the separation of
two diverging trajectories that start simultaneously at the same initial spatial
point, but
have slightly different initial velocity directions, specified by
angles $\theta$ and $\theta^{\prime}$, with $\theta \left(
t=0\right) =\theta _0$, and
$\theta ^{\prime }\left( t=0\right)
^{}=\theta _0+\delta \theta _0$. We choose to measure the separation of
trajectories along a line which is at all times $t>0$ 
perpendicular to the
reference trajectory. Because the trajectories are curved by the
thermostatted field, the intersection of this line with the
adjacent trajectory does not occur at the point that a particle
following the adjacent trajectory will reach at time $t$, even to
first order in the infinitesimal quantities, such as $\delta \theta_0$. 
In order to determine the Lyapunov exponents, one must take
effects produced by the curving of the trajectories into account (see
Fig. 1).
The separation between the two trajectories at 
time $t$, $\delta S\left( t\right) $, can be written
as the product of a radius of curvature $\rho (t)$
and $\delta^{\prime}\theta \left(t \right)$, the difference between the 
velocity angles of the two trajectories at
the points of intersection with the 
perpendicular line, or 
$\rho \left( t\right)=S(t)/ \delta ^{\prime }\theta \left( t\right).$ 
 The radius of 
curvature changes continuously
between collisions, and instantaneously 
at the collisions of the particle with the scatterers. Simple geometric
considerations yield an equation for the rate of change of $\rho(t)$
between collisions
as
\begin{equation}
\dot{\rho}=v+(\rho \varepsilon) \cos\theta+\frac{\rho^2 
\varepsilon^2}{v} \sin ^2\theta.
\label{3}
\end{equation}
The instantaneous change in the radius of
curvature due to a collision is given by
\cite{sinai}  
\begin{equation}
\frac 1{\rho _{+}}=\frac 1{\rho _{-}}+\frac 2{a\cos \phi },  \label{4}
\end{equation}
where $\vec{v}\cdot \hat{k}=-\cos \phi$, $\phi$ is the angle of
incidence at collision, and $\rho _{-}$ and $\rho _{+}$ are the 
radii of curvature 
before and after collision, respectively.

We now compute the positive Lyapunov exponent $\lambda_+$
from the rate of separation of diverging trajectories. The result of 
Sinai \cite{sinai} for the positive
Lyapunov exponent in a field-free Lorentz gas can be straightforwardly generalized to include
the case under discussion:
\begin{equation}
\lambda _{+}=\lim_{T\rightarrow \infty }\frac 1T\int_{t_0}^{t_0+T}
\frac{v}{\rho(t)}\,dt=\left\langle \frac v\rho \right\rangle _{s.s.}.  \label{5}
\end{equation}
Here we assume ergodicity and calculate the ensemble average in the
steady state, denoted by
the angular brackets with subscripts {\it s.s.}, using
a non-equilibrium steady state distribution function for the moving
particle in the constant thermostatted applied field. The main idea
is to consider an ensemble of similarly prepared systems, assume that
the distribution function for the moving particle, in this ensemble, reaches a spatially
homogeneous steady state (since there is no way to distinguish one
spatial point from another in the ensemble average), and to derive and
solve a Lorentz-Boltzmann
(LB) equation for the distribution function for the moving
particle $f\left(\vec{v},\rho \right)$, where the variables
include both the velocity and the radius of
curvature describing the separation of trajectories of the moving
particle and an adjacent trajectory, as described above.

Thus we write 
\begin{equation}
\lambda _{+}=\int d\vec{v}\int_0^\infty d\rho \frac v\rho
f\left( \vec{v},\rho \right) ,  \label{6}
\end{equation}
assuming that $f$ has been normalized to unity. The LB
equation for $f$ is
\begin{equation}
\nabla _{\vec{v}} \cdot \left( f\dot{\vec{v}}\right) +\frac \partial {\partial
\rho }\left( \dot{\rho}f\right) =\left( \frac{\partial f}{\partial t}\right)
_{coll}.  \label{7}
\end{equation}
Since the dynamics between collisions is not
Hamiltonian, the usual form of the streaming terms on the left hand
side of the LB equation must be replaced by the form that reflects the
total conservation of particles under the actual, non-Hamiltonian
dynamics. The left hand side of the LB equation can be obtained from
the equations of motion, Eq. (\ref{2}), and Eq. (\ref{3}) for $\dot{\rho}$
\begin{equation}
-\varepsilon \frac \partial{\partial \theta}(f\sin \theta)+\frac
\partial{\partial \rho} \left[ \left(v+\rho\varepsilon \cos\theta
+\frac{\rho^{2}\varepsilon^{2}}{v} \right)f \right]=\left(
\frac{\partial f}{\partial t} \right)_{coll}.
\label{8}
\end{equation}
where we used the constant speed of the particle to denote
$f(\vec{v},\rho)$ by $f(\theta,\rho)$.
The right hand side of the LB equation is the change in $f$ due to
collisions given previously \cite{vbd95} as
\begin{eqnarray}
\left( \frac{\partial f}{\partial t}\right)_{coll}&=& nav \int_{- \pi/2}^{
\pi/2} d \phi \cos \phi\int_{0}^{\infty}d \rho' \delta \left( \rho -
\frac{(a \cos \phi)/2}{1+(a \cos \phi)/2 \rho'}
\right) f(\vec{v'},\rho')  \nonumber \\ 
&-& 2navf(\vec{v},\rho).
\label{9}
\end{eqnarray} 
To leading order in the density, the collision terms can be simplified
further by approximating the factor $1+(a \cos \phi)/2 \rho'$ inside
the $\delta$ function by unity since $\rho'$ is typically on the order
of the mean free path, so that $\rho' \gg a$. Then Eq. (\ref{8}) can be
solved by expanding $f$ as a power series in $\varepsilon$ and
inserting the solution into Eq. (\ref{6}). We obtain
\begin{equation}
\lambda_{+}(\varepsilon)=\lambda_{0}-\frac{11}{48}t_{0}\varepsilon^2+O(\varepsilon^4),
\label{10}
\end{equation}
where $t_0=\ell/v,$ with $\ell=(2na)^{-1}$ the mean free path of the
moving particle, and $\lambda_0$ is given in Ref.
\cite{vbd95}. 

The calculation of $\lambda_{-}(\varepsilon)$ requires the study of
trajectories that {\em asymptotically} converge to the  reference
trajectory of the moving particle rather than diverge from it. That
is, the negative Lyapunov exponent can only be determined if one can
find trajectories that lie on the stable manifold of the reference
trajectory, which is typically difficult since almost all of the
adjacent trajectories will eventually diverge from it. To overcome
this difficulty we consider the time reversed motion of the moving
particle \cite{timerev}. This allows us to consider diverging
trajectories again. We thus consider the steady state distribution
function $f_{-}(\vec{v},\rho)$ for trajectories of the particle with
velocity $\vec{v}$ and radius of curvature $\rho$ on the time-reversal
of the stable manifold. The equation for $f_-$ takes the form of an
anti-Lorentz-Boltzmann (ALB) equation. This unusual form is
dictated by the observation that if the moving particle and the
scatterer with which it collides are uncorrelated {\em before}
collision in the forward motion, then in the time reversed motion, the
moving particle and scatterer will be uncorrelated {\em after} the
collision. That is, to obtain the ALB, one must use the {\em
Stosszahlansatz} for the {\em exiting} collision cylinders, rather than
for those before the collision \cite{lvbd,cober}. The ALB then reads
\begin{equation}
-\varepsilon \frac \partial {\partial \theta }(f_{-}\sin
 \theta )+\frac
\partial {\partial \rho }\left[\left(v+\rho \varepsilon \cos \theta +
 \frac{\rho^2 \varepsilon^2}{v}\sin ^2\theta \right)f_{-}\right]=
  \left( \frac{\partial f_{-}}{\partial t}
\right) _{coll}  \label{11}
\end{equation}
where
\begin{eqnarray}
\left( \frac{\partial f_{-}}{\partial t}\right) _{coll} &=&nav\int_{-\pi
/2}^{\pi /2}d\phi \cos \phi \int_0^\infty
d\rho ^{\prime }\, \delta (\rho -(a\cos \phi)/2 )f_{-}(\vec{v},\rho ^{\prime })  \nonumber \\
&&-nav\int_{-\pi /2}^{\pi /2}d\phi \cos \phi \frac{\int_0^\infty d\rho
^{\prime }f_{-}\left( \vec{v'},\rho ^{\prime }\right) }{%
\int_0^\infty d\rho ^{\prime }f_{-}\left( \vec{v},\rho^{\prime }\right) }%
f_{-}\left( \vec{v},\rho \right) .    \label{12}
\end{eqnarray}

The first term on the r.h.s. of Eq. (\ref{12}) is the gain term. It is
constructed by: (a) using the {\it Stosszahlansatz} for the collision
cylinders centered on particles which are produced in collisions with
velocity $\vec{v}$, (b) requiring that the radius of curvature after
collision be $\rho$, and (c) noting that before collision, the radius
of curvature will typically be on the order of a mean free path, so that after
collision $\rho$ will be very close to $(a/2)\cos \theta$.
The loss term in Eq. (\ref{12}) can be obtained by noting that: (a)
The rate at which particles with velocity $\vec {v}$ disappear due to
collisions with angle of incidence $\phi$ is $nav \cos \phi
\int_{0}^{\infty}d \rho' f_{-}(\vec{v'},\rho')$, and (b) the fraction
of those which disappear having radius of curvature $\rho$ is
$f_{-}(\vec{v},\rho)[\int_{0}^{\infty}d \rho'
f_{-}(\vec{v},\rho')]^{-1}$. These considerations lead directly to the
r.h.s. of Eq. (\ref{12}), and a more detailed explanation will be given
elsewhere \cite{lvbd}.

Eq. (\ref{11}) can be
solved by imposing the requirement that
\begin{equation}
\int d\rho f_{-}\left( \vec{v},\rho \right)= f_{-}(\vec{v}) \label{13}
\end{equation}
with $ f_{-}(\vec{v}) $ the time reversed steady
state solution of the  spatially homogeneous  Lorentz-Boltzmann
equation. To understand this condition, we note that when Eq. (\ref {11}) 
is integrated over all values of $\rho $ to obtain an equation for 
the velocity distribution function 
$ f_{-}\left( \vec{v}\right) $ one obtains an ALB equation
with a collision operator that has the opposite sign from the usual
Lorentz-Boltzmann equation, due to the fact that
we are considering the time-reversed motion, and using the {\it Stosszahlansatz}
after collisions, \cite{cober}. For consistency, we then require that the
steady state solutions of the $\rho$ integrated LB and ALB equations
be related by a simple time-reversal operation, as indicated by Eq. (\ref{13}).

The negative Lyapunov exponent is obtained from $f_{-}$ by
\begin{equation}
\lambda _{-}=-\int d\vec{v}\int_0^\infty d\rho \frac v\rho f_{-}\left( \vec{v%
},\rho \right)   \label{14}
\end{equation}
with $f_{-}$ normalized to unity. The negative sign on the r.h.s. of
Eq. (\ref{14}) is the result of the time reversal procedure. Again, the equation for $f_{-}$ can be
solved by expanding $f_{-}$ in powers of $\varepsilon.$ 
We finally obtain
\begin{equation}
\lambda _{-}=-\lambda _0-\frac 7{48}t_0 \varepsilon^2 +O\left(
\varepsilon^4\right) .  \label{15}
\end{equation}
Using the relation between the diffusion coefficient $D$ and the sum of
the Lyapunov exponents discussed earlier, we recover the correct low
density value $D=(3/8)t_0 v^2.$ 

      It is important to 
compare these results with those obtained from computer
simulations of the same system. For this purpose we distributed $10^5$
non-overlapping scatterers randomly in a square simulation cell with
periodic boundaries. Between 
collisions the trajectory of the moving
particle is computed from an analytical solution \cite{mh87} of the
thermostatted equations of motion. The collision points with the
scatterers are determined numerically with an accuracy of
$10^{-12}$. The Lyapunov exponents were computed in tangent space by a
simultaneous integration of the linearized equations of motion for the
intercollisional streaming and an exact linearization of the map which relates the separation of trajectories
after collision to that before collision \cite{dgph}.
In Fig. 2 we show the
deviation of the non-vanishing Lyapunov exponents from their
equilibrium values as a function of the squared applied field. All numbers are
made dimensionless by using $a,v,m$ and $q$ as the respective units for length
velocity, mass, and charge.
Two reduced
scatterer densities $n^* = na^2$ were considered, 0.001 (circles) and
0.002 (squares). Each point is obtained by averaging over 10
simulation runs with different scatterer configurations and a total of
$4 \times 10^6$ collisions for each run. 
The standard deviation  for the exponents
is $\sim 0.1 \% $. The lines refer to the theoretical expressions,
Eq. (\ref{10}) and Eq. (\ref{15}). Both
$\lambda_+$ and $\lambda_-$ exhibit the predicted quadratic weak field
behavior. These results
are clearly consistent with the theoretical predictions for the
field-dependent Lyapunov exponents, though at densities somewhat
higher than those for which the LB results hold
without density corrections.
Fig. 3 shows $D/ (\ell v)$ as a function of $\varepsilon^2 \ell a/v^2$ for
$n^* =0.001$ (full circles) and 0.002 (open circles).
The diffusion
coefficient $D$ is obtained from the conductivity $\sigma$ through
the Einstein formula, and $\sigma$ is numerically computed
from $ \sigma = \langle qv_x\rangle_{s.s.}/E$, the ratio of the
time-averaged dissipative current to the applied field.
Alternatively, $D$ can be obtained also from the Lyapunov exponents as
indicated above. On the scale of Fig. 3 both methods yield indistinguishable
results.

     We mention here that to obtain agreement between the simulation 
data for $D$ and theory it was necessary to add to the low-density
Lorentz-Boltzmann result $(3/8) \ell v$, the dotted top
horizontal line in Fig. 3, 
known \cite{bruin} moderate density corrections, 
which include essential contributions from
logarithmic terms of the form $n^{*} \ln n^{*}+O(n^*)$. The numerical nonequilibrium results
converge well to these corrected values for vanishing field indicated
by the dashed ($n^* = 0.001$) and solid ($n^* =0.002$) horizontal
lines. Thus the computer results confirm the presence of these
non-analytic terms in the density expansion.  

       As a check we computed the Lyapunov exponents for still smaller
densities, $n^* = 0.0001$, and fields $ < 10^{-5} $ with a very efficient
direct-simulation Monte-Carlo method \cite{delhap96} and found perfect
agreement with our other simulation results quoted above.

We conclude by remarking that this
work illustrates the power of kinetic theory methods for computing
quantities of interest to both statistical mechanics and dynamical
systems theory \cite{vbd95,lvbd,egdc}.

J.R.D. thanks A. Latz for helpful discussions, and the National
Science foundation for Support under Grant
No. PHY-93-21312. H.A.P. and Ch.D. gratefully acknowledge support
from the Fonds zur F\"orderung der wissenschaftlichen Forschung, Grant
No. P09677, and the generous allocation of computer resources by the
Computer Center of the University of Vienna. E.G.D.C. is
indebted to the U. S. Department of Energy for support under grant
DE-FG 02-88-ER 13847.

\newpage
\begin{center}
      {\bf Figure captions}
\end{center}

Fig. 1: Geometry for the curvature $\rho(t)$ in the applied field.  

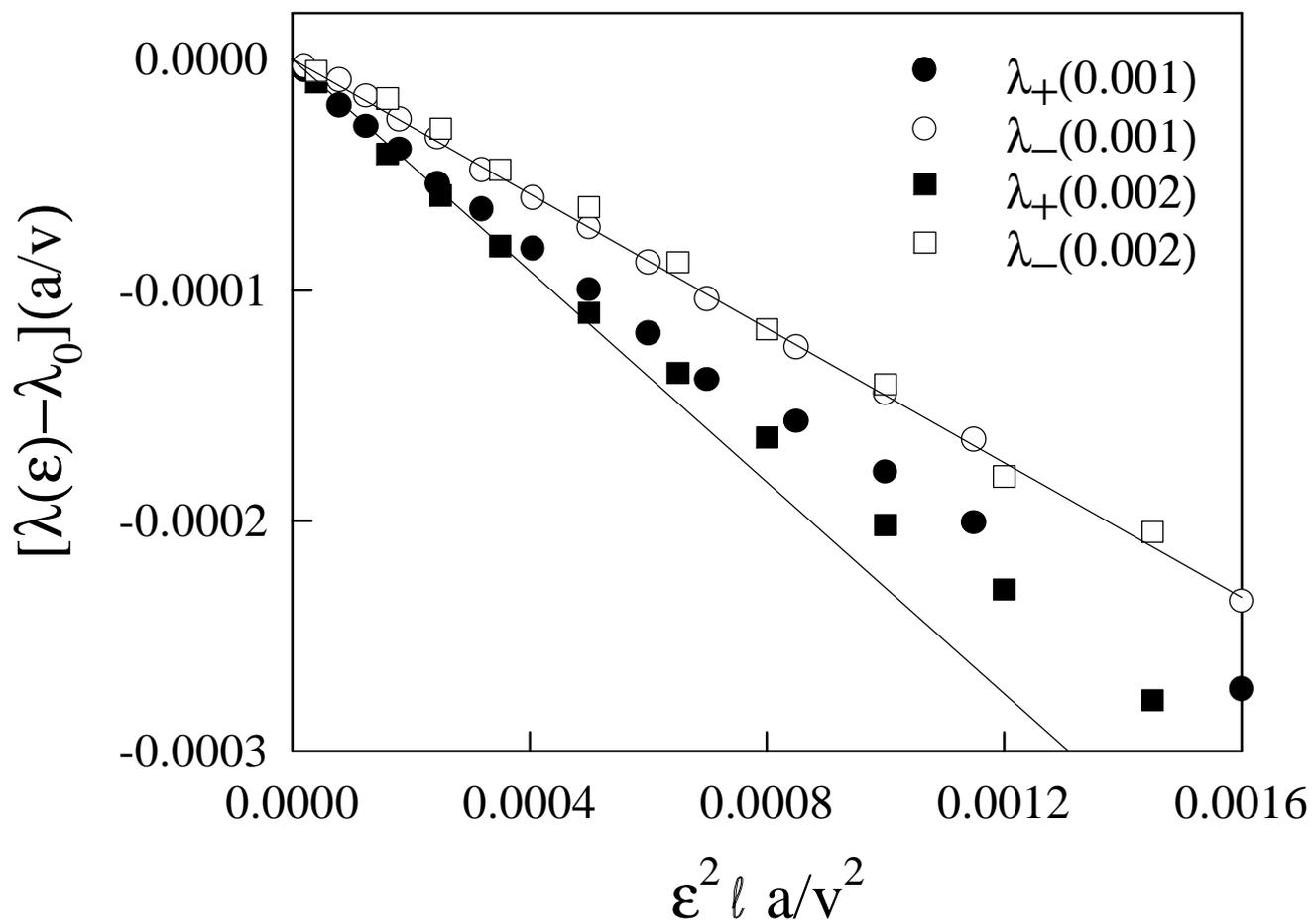
Fig. 2: Field dependence of the deviation of the Lyapunov exponents from their
        equilibrium value for the reduced densities
        $n^* = 0.001$ and $n^* = 0.002$.
        $\varepsilon = qE/vm$, where $E$ is the applied field.
        $\ell$ is the mean free path, and $q,m$, and $v$ are
        the charge, mass, and velocity of the particle, respectively.
        $a$ is the diameter of the scatterers.

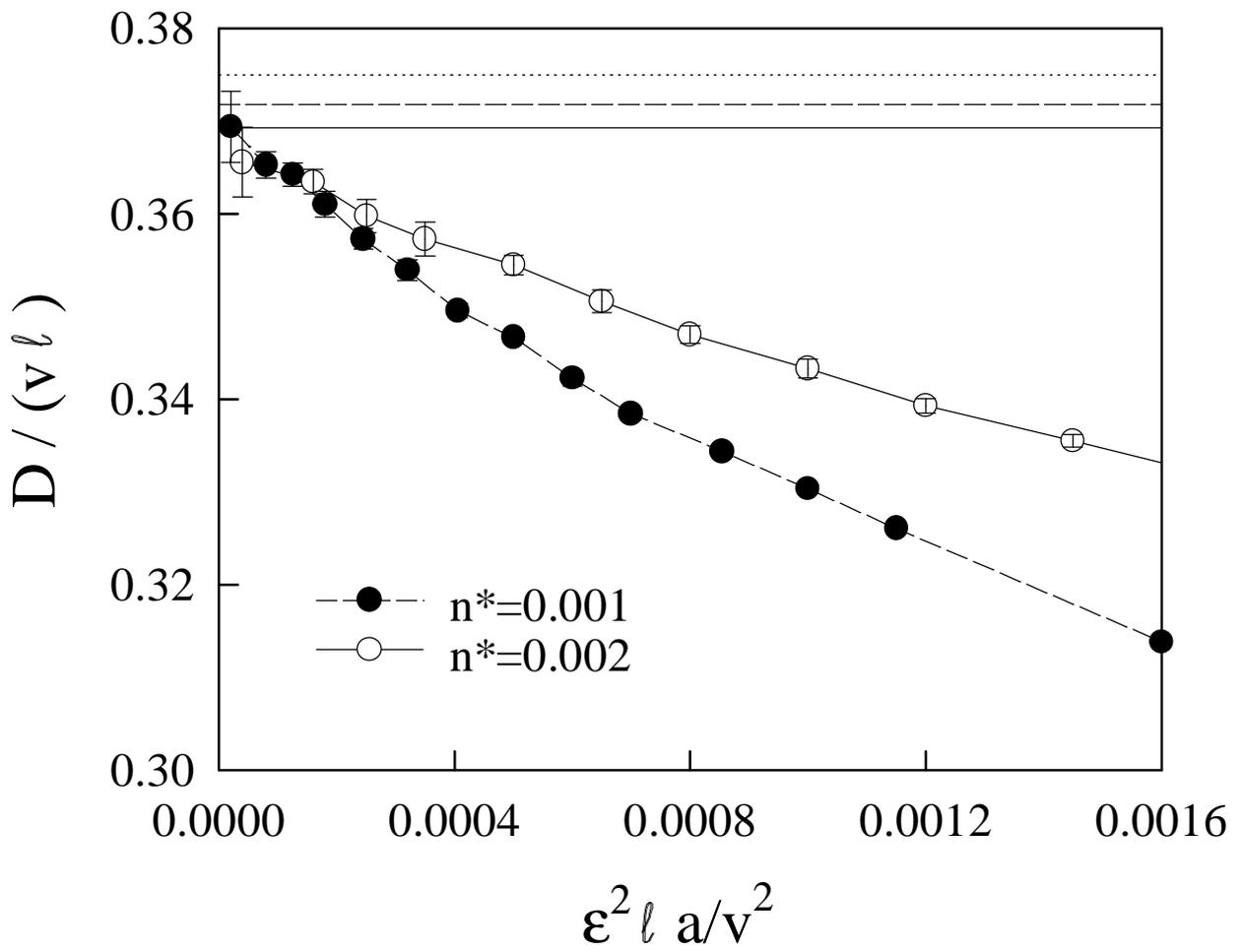
Fig. 3: Dependence of the diffusion coefficient $D$ on the field
        for the two densities $n^* = 0.001$ and $0.002$.
        All other quantities are as in Fig. 2. The horizontal
        lines are explained in the text.

\begin{references}
\bibitem{timerev} Note the same is done in computer simulations
designed for measuring $\lambda_{-}$, cf. Refs.(2,3) above.
\end{references}
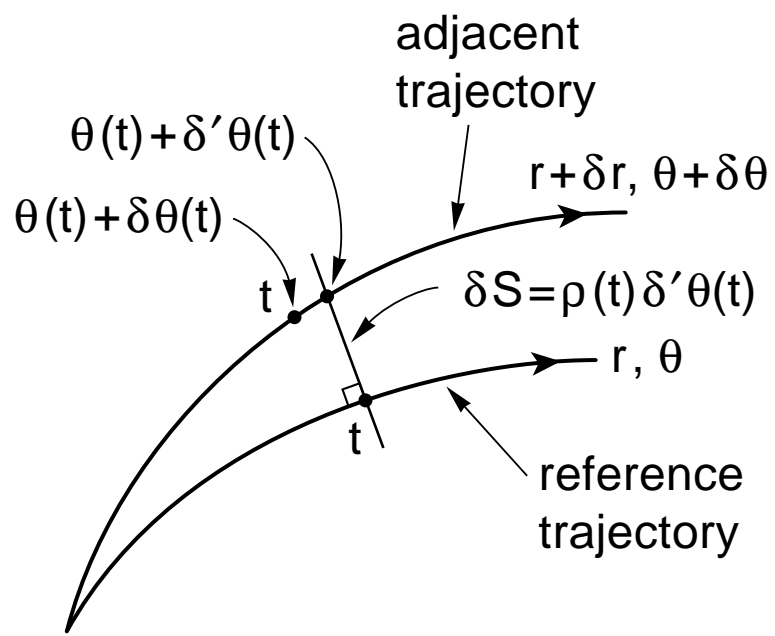
\end{document}